\documentclass[10pt, a4paper, twocolumn]{article} 

\usepackage[english]{babel} 

\usepackage{microtype} 

\usepackage{amsmath,amsfonts,amsthm} 

\usepackage[svgnames]{xcolor} 

\usepackage[hang, small, labelfont=bf, up, textfont=it]{caption} 

\usepackage{booktabs} 

\usepackage{lastpage} 

\usepackage{graphicx} 

\usepackage{enumitem} 
\setlist{noitemsep} 

\usepackage{sectsty} 
\allsectionsfont{\usefont{OT1}{phv}{b}{n}} 

\usepackage{geometry} 

\usepackage[T1]{fontenc} 
\usepackage[utf8]{inputenc} 

\usepackage{XCharter} 

\usepackage{fancyhdr} 
\fancypagestyle{firstpage}{ 
	\fancyhf{}
}

\newcommand{\authorstyle}[1]{{\large\usefont{OT1}{phv}{b}{n}\color{DarkRed}#1}} 

\newcommand{\institution}[1]{{\footnotesize\usefont{OT1}{phv}{m}{sl}\color{Black}#1}} 

\usepackage{titling} 

\newcommand{\HorRule}{\color{DarkGoldenrod}\rule{\linewidth}{1pt}} 

\pretitle{
	\vspace{-50pt} 
	\HorRule\vspace{10pt} 
	\fontsize{16}{36}\usefont{OT1}{phv}{b}{n}\selectfont 
	\color{DarkRed} 
}

\posttitle{\par\vskip 15pt} 

\preauthor{} 

\postauthor{ 
	\vspace{10pt} 
	\par\HorRule 
	\vspace{-32pt} 
}

\usepackage{lettrine} 
\usepackage{fix-cm}	

\newcommand{\initial}[1]{ 
	\lettrine[lines=3,findent=4pt,nindent=0pt]{
		\color{DarkGoldenrod}
		{#1}
	}{}%
}

\usepackage{xstring} 

\newcommand{\lettrineabstract}[1]{
	\StrLeft{#1}{1}[\firstletter] 
	\initial{\firstletter}\textbf{\StrGobbleLeft{#1}{1}} 
}

\usepackage{siunitx}
\usepackage[version=4]{mhchem}

\title{A symmetry-derived mechanism for atomic resolution imaging} 

\author{
	\authorstyle{Matus Krajnak\textsuperscript{1,2} and Joanne Etheridge\textsuperscript{1,3}} 
	\newline\newline 
	\textsuperscript{1}\institution{Department of Materials Science and Engineering, Monash University, Victoria 3800, Australia}\\ 
	\textsuperscript{2}\institution{Current address: Quantum Detectors Ltd, R104 RAL, Harwell Oxford OX11 0QX, UK}\\ 
	\textsuperscript{3}\institution{Monash Centre for Electron Microscopy, Monash University, Victoria 3800, Australia} 
}

\date{} 

\begin{document}

\maketitle 

\thispagestyle{firstpage} 


\lettrineabstract{We introduce a new image contrast mechanism for scanning transmission electron microscopy (STEM) that derives from the local symmetry within the specimen. For a given position of the electron probe on the specimen, the image intensity is determined by the degree of similarity between the exit electron intensity distribution and a chosen symmetry operation applied to that distribution. The contrast mechanism detects both light and heavy atomic columns and is robust with respect to specimen thickness, electron probe energy and defocus. Atomic columns appear as sharp peaks that can be significantly narrower than for STEM images using conventional disc and annular detectors. This fundamentally different contrast mechanism complements conventional imaging modes and can be acquired simultaneously with them, expanding the power of STEM for materials characterisation.}


\begin{figure*}[h]
	\includegraphics[width=\textwidth]{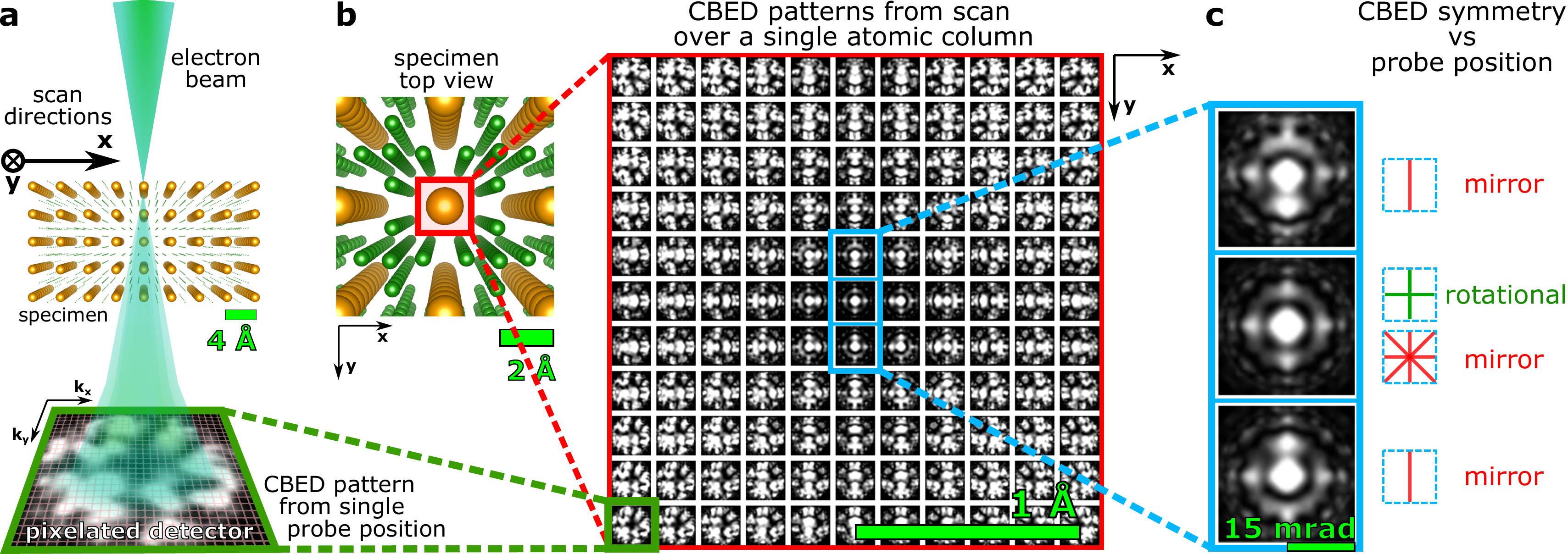}
	\caption{Schematic of a Symmetry STEM experiment. \textbf{a} An atomic-scale electron probe scans the specimen. A convergent beam electron diffraction pattern is imaged in the far field by a fast pixelated detector (green box) for each point of the scan. \textbf{b} Grid of simulated CBED patterns resulting from a scan across a \ce{Ce} atomic column (red box) in a \ce{CeB6} crystal. Each CBED pattern is arranged according to the position of the electron probe in real space, scan step size is $20$ picometres. \textbf{c} CBED patterns for the probe on the centre of the \ce{Ce} column (middle figure) and shifted $20$ picometres either side of the centre (but still on the atomic column) (top and bottom figures), showing the rapid change in pattern symmetry (4mm to m). \label{fig1}}
\end{figure*}


In scanning transmission electron microscopy (STEM), images can be generated by scanning an electron beam across the object and, at each position of the electron beam, detecting the scattered electron intensity distribution after transmission through the specimen. The most common STEM imaging modes integrate the scattered intensity in the diffraction plane across a particular angular range, using either disc or annular detector geometries, to generate phase-contrast \textit{Bright Field} (BF) and adsorptive-contrast \textit{High Angle Annular Dark Field} (HAADF) images, respectively, (or a mix of phase and adsorptive contrast in the case of \textit{Annular Bright Field} (ABF)). Recently, the advent of fast read-out, high dynamic range detectors \cite{battaglia2010characterisation,ballabriga2011medipix3,ryll2014pnccd,tate2016high,mir2017characterisation,faruqi2018direct} has enabled the full angular distribution of scattered intensity to be recorded at each beam position. This represents a revolution in STEM, providing access to a vast and rich palate of additional specimen information \cite{ophus2019four}. Fast detectors have already been applied, for example, to improve STEM spatial resolution using ptychography \cite{pennycook2015efficient,yang2015efficient,yang2016simultaneous,jesse2016big,jiang2018electron}, to map electric \cite{muller2017measurement,hachtel2018sub} and magnetic fields \cite{krajnak2016pixelated,tate2016high,mcgrouther2016internal,mcvitie2018transmission,nord2019strain}, strain \cite{jesse2016big,pekin2017optimizing,grieb2018strain,han2018strain,nord2019strain}, polarization domains \cite{nguyen2016reconstruction}, and octahedral tilts \cite{nord20183d}, representing just the beginning of this powerful new era in STEM.

Here we propose a new image contrast mechanism for atomic resolution STEM based on a measurement of the degree of symmetry in the scattered intensity distribution at each point of a scan: Symmetry STEM (S-STEM). By virtue of the strong electron-specimen interaction and resultant dynamical scattering, the symmetry of the illuminated specimen volume is encoded in the symmetry of the scattered intensity distribution, independent of the specimen thickness and accelerating voltage \cite{moodie2010dynamical,cowley2006electron,buxton1976symmetry,tanaka2010point}. In this paper, the scattered intensity distribution is in the form of a convergent beam electron diffraction (CBED) pattern (Fig. \ref{fig1}a), the most common case for STEM \cite{spence1978lattice}, but the approach can be applied in principle to any form of scattered intensity distribution in any optical plane and also to other scanning microscopy techniques.

The `degree' of symmetry in a pattern can be analysed by a comparison of the scattered intensity  distribution with itself after an applied symmetry operation \cite{masuda1993detection}. For a given two dimensional (2D) pattern $\mathbf{A}$, the symmetry intensity $I$ is given by:

\begin{equation}
I = \textrm{max} \left[ \mathbf{A} \ast  \, _{\:\!\:\! {operation}}^{{symmetry}}\left(\mathbf{A}\right)\right], \label{eq:image_contrast} 
\end{equation}
where $\ast$ is a normalised cross-correlation and the symmetry operation can be chosen (for example, a rotation or a mirror). If $\mathbf{A}$ is invariant under the symmetry operation, then the intensity will be maximum, $I=1$, and $I<1$ if the symmetry is not matched. Analogous algorithms have been applied to visualise biological macromolecules \cite{crowther1971harmonic,frank2006three} and in measurements of local polarization domains \cite{tsuda2017nanometer,shao2017lattice}. Here, our goal is different, namely, to deliver an image contrast mechanism that derives from spatial variations in symmetry, measured at picometre intervals.

In the case of Symmetry STEM, the intensity will be calculated from each CBED pattern at each point of a 2D scan, $(x,y)$, which can be plotted as an image $\mathbf{I} = I_{x,y}$. Each CBED pattern resolves the electron distribution in reciprocal space at a particular point, $(x,y)$, of a scan $\mathbf{A} = \mathbf{A}_{x,y}(k_x,k_y)$, denoting one point in a so called four dimensional (4D) STEM dataset \cite{ophus2014recording}. Data processing was based on methodology introduced in Refs. \cite{krajnak2016pixelated,krajnak2017advanced,krajnak2018git} and implemented in GPU accelerated ArrayFire library \cite{Yalamanchili2015}. 
The application of Equation \ref{eq:image_contrast} to 4D-STEM dataset generates an entirely different image contrast mechanism, neither phase-contrast nor adsorption-contrast, which provides access to new specimen information at the atomic level.

\begin{figure}[b!]
	\includegraphics[width=\columnwidth]{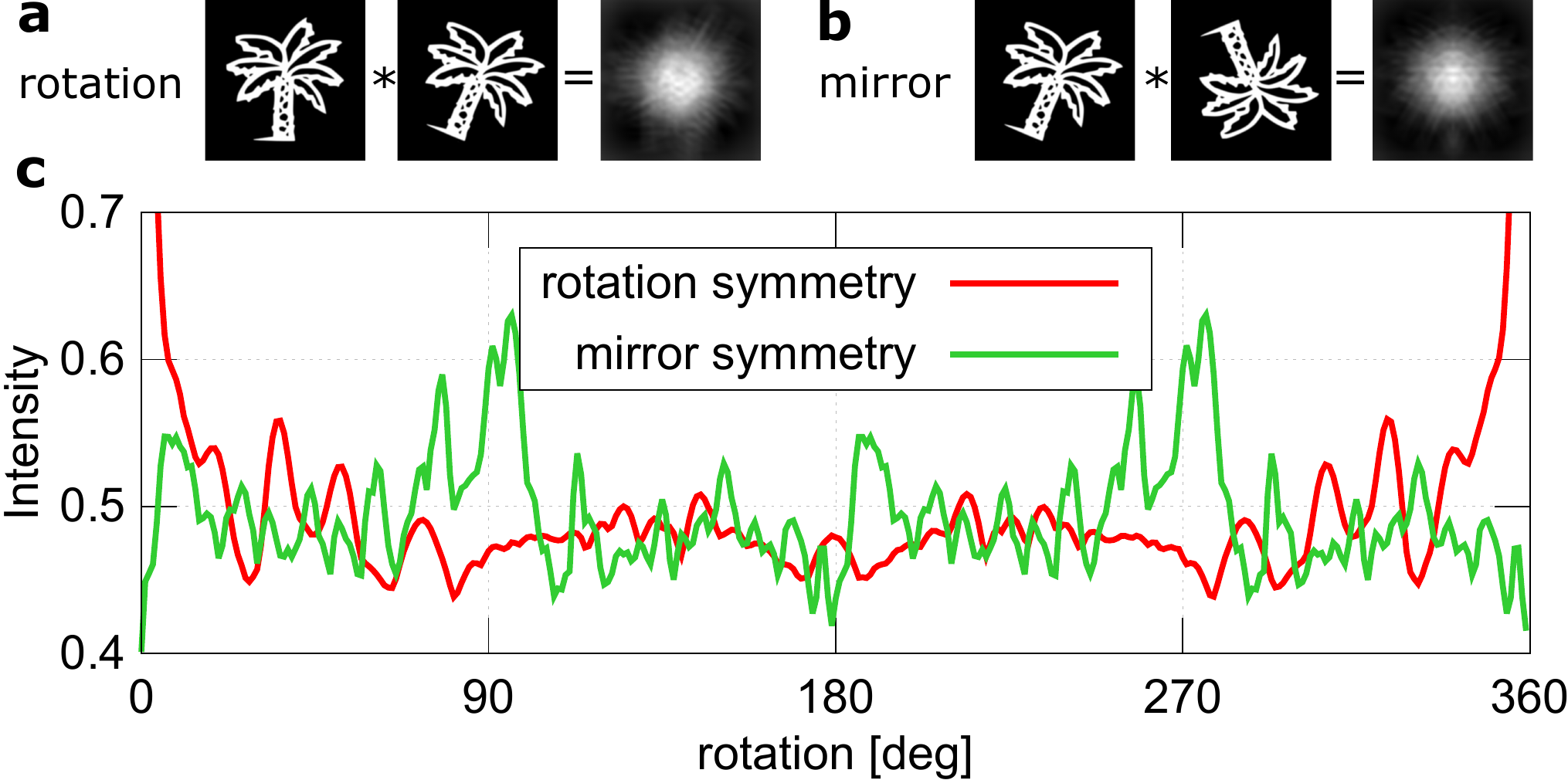}
	\caption{Example of symmetry search for a palm tree pattern. \textbf{a} and \textbf{b} show cross-correlation patterns for rotation and mirror symmetry, respectively. The pattern was rotated by $20^{\circ}$. Mirror symmetry was tested along the vertical axis of the pattern after the rotation. \textbf{c} Cross-correlation plot for $0^\circ$ to $360^\circ$ rotation. $I$ is only shown for the range containing maximum variations. \label{fig2}}
\end{figure}

\begin{figure*}[t]
	\includegraphics[width=\textwidth]{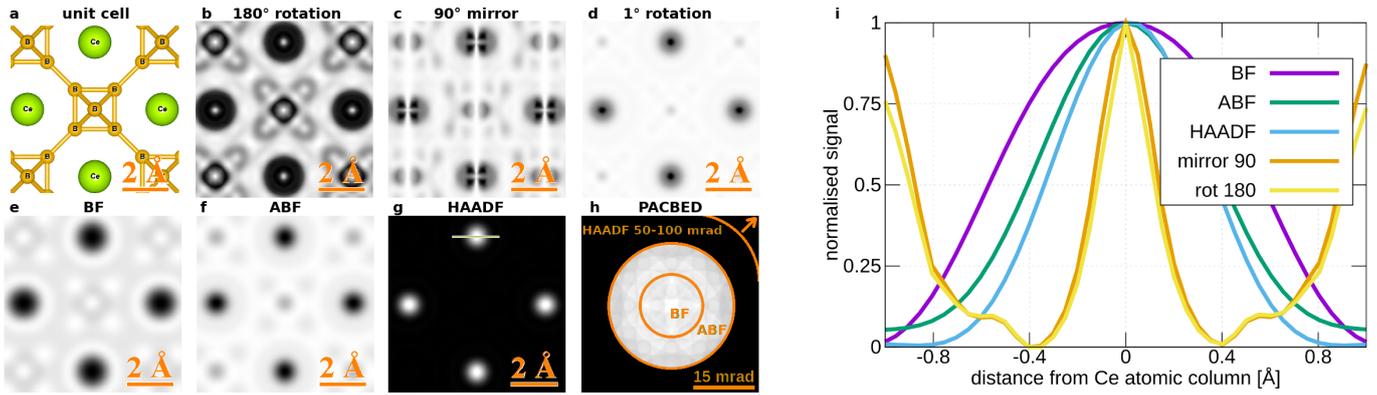}
	\caption{\textbf{a} Schematic of \ce{CeB6} unit cell in $\left<100\right>$ orientation. \textbf{b}-\textbf{d} Reconstructions of Symmetry STEM images for different applied symmetry operations.  \textbf{e}-\textbf{g} Reconstructions of conventional STEM images with integration areas shown on the position averaged diffraction pattern in \textbf{h}. \textbf{i} Comparison of the corresponding \ce{Ce} image peak width for S-STEM {b} and {c} and conventional images {e}-{g} (line in image {g} shows where the profiles were taken). \label{fig3}}
\end{figure*}


\section*{Results}
\subsection*{Calculations}
The sensitivity to local symmetry that underpins the Symmetry STEM contrast mechanism is illustrated with a STEM simulation on a test case, \ce{CeB6}. \ce{CeB6} has both small (B-B : $\SI{1.2}{\angstrom}$), and large (Ce-Ce : $\SI{4.1}{\angstrom}$) column spacings and comprises light (B = 5) and  heavy (Ce = 58) atoms, which are challenging to image simultaneously in BF or ADF STEM. Figure \ref{fig1}b shows an array of simulated CBED patterns corresponding to a $\SI{0.5}{\angstrom}$ FWHM probe scanning across the \ce{Ce} column in $\left<100\right>$ oriented \ce{CeB6}. Each simulated CBED pattern is arranged according to the corresponding real space position, $(x,y)$, of the probe, with the \ce{Ce} column position centred at the centre of the array. The GPU accelerated parallel implementation of the multislice simulations were performed using the Prismatic software package \cite{pryor2017streaming,orphus2017afast,kirkland2010advanced} using parameters described in Methods - STEM simulation. The arranged CBED patterns give a clear sense of how the symmetry changes as the probe is scanned across the \ce{Ce} atomic column in $\SI{0.2}{\angstrom}$ steps (Fig. \ref{fig1}c).  For example, there is an immediate shift from 4-fold and multiple mirrors (4mm) to a single mirror symmetry (m) as the probe centre moves from the absolute centre of the atomic column to just $20$ picometres off-centre (but is nevertheless still located on the atomic column). This highlights the acute sensitivity to local specimen symmetry that is delivered by dynamical scattering \cite{cowley2006electron,buxton1976symmetry,tanaka2010point,kimoto2011spatially} and forms the basis of image contrast in Symmetry STEM.

Equation \ref{eq:image_contrast} provides an extremely efficient method for distilling the local symmetry information present in the pattern. This is illustrated generically in Fig. \ref{fig2} for an intensity distribution, $\mathbf{A_p}$, in the form of a palm tree. Two classes of symmetry are tested, a rotation and mirror symmetry. These symmetry operations can be tested for an arbitrary angle on this arbitrary pattern, $\mathbf{A_p}$ (e.g. Figs. \ref{fig2}a and \ref{fig2}b, test a $20^\circ$ rotation for the palm tree pattern). The cross-correlation is calculated for $0^\circ$ to $360^\circ$ (Fig. \ref{fig2}c). It can be seen that even if the pattern does not possess a perfect symmetry element, there still exist local maxima in the analysis. For the palm tree pattern this signal arises when the leaves are overlapping after application of a given symmetry operation. It should be noted that only $0^\circ$ to $180^\circ$ angles need to be calculated because only the relative rotation matters. $1^{\circ}$ clockwise and $1^{\circ}$ anti-clockwise rotations have the same maximum of cross-correlation (ignoring interpolation errors). For additional notes about this symmetry analysis please see Supplementary information.

To demonstrate Symmetry STEM, the symmetry analysis of Eq. \ref{eq:image_contrast} will now be applied to the simulated scanning CBED dataset (Fig. \ref{fig1}) across a field of view slightly larger than a \ce{CeB6} unit cell (Fig. \ref{fig3}a), again using the parameters described in Methods - STEM simulation. Symmetry STEM images corresponding to $180^{\circ}$ rotation, $90^{\circ}$ mirror and $1^{\circ}$ rotation are generated (Figs.~\ref{fig3}b-d, respectively).  For comparison, standard STEM BF, ABF and HAADF images were  also reconstructed (Figs. \ref{fig3}e-g, respectively) by integrating across the angular ranges indicated in (Fig. \ref{fig3}h) for each probe position.

As anticipated, the S-STEM images exhibit atomic-scale contrast, revealing local maxima wherever some degree of the applied symmetry element is present, reaching a maximum value near one when there is an identity, such as on the \ce{Ce} column at $180^{\circ}$ rotation and $90^{\circ}$ mirror. $180^{\circ}$ rotation symmetry shows exceptionally intense and sharp contrast for \ce{Ce}, small local maxima at all \ce{B} positions, and also it highlights the $180^{\circ}$ symmetry with a broad maxima half way between \ce{Ce} atomic columns along $\left<100\right>$ (Fig. \ref{fig3}b). $90^{\circ}$ mirror symmetry shows bright contrast where mirror planes within the lattice match the symmetry (Fig. \ref{fig3}c). An interesting contrast arises when a small rotation symmetry of $1^{\circ}$ is measured (Fig. \ref{fig3}d) giving extra sensitivity to the rate of change of the local symmetry. Strong `intensity' appears where the CBED pattern is rotationally symmetric or it varies slowly with angle $(k_x, k_y)$, namely at the position of asymmetric \ce{B} atom sites, with weak intensity at the \ce{Ce} sites where the pattern varies rapidly.

Importantly, local maxima at atomic sites in the Symmetry STEM images are exceptionally sharp with low intensity `moats' around them. This is particularly the case for the peaks at \ce{Ce} columns with a FWHM of $\sim$\SI{0.25}{\angstrom} (Fig. \ref{fig3}i), which are significantly narrower than the peaks in the corresponding conventional BF, ABF and HAADF STEM images (using a diffraction limited probe with FWHM \SI{0.8}{\angstrom} (Fig. \ref{fig3}i)). Local maxima due to the presence of a symmetry element but in the absence of an atomic column do not show the `moat' because the rate of change of specimen symmetry is more slowly varying than in the presence of an atomic column. This enables atom sites to be distinguished from atom-free symmetry sites. This distinction can be further checked when S-STEM images derived from different symmetry elements are compared.

\begin{figure}[b!]
	\includegraphics[width=\columnwidth]{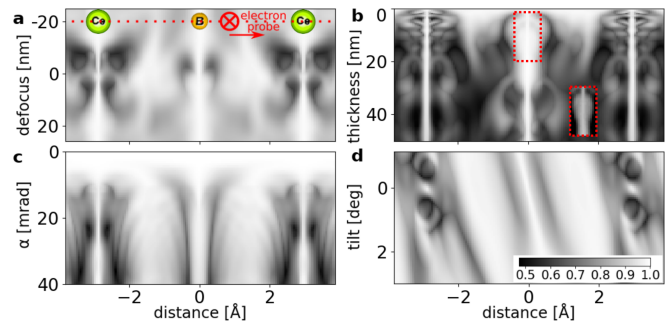}
	\caption{Symmetry STEM intensity for a scan across \ce{Ce-B-Ce} atoms along $\langle100\rangle\>$. \textbf{a} Intensity versus defocus (varied between \SI{-25}{nm} to \SI{27}{nm} with the sample located at $0$--\SI{3}{nm}). \textbf{b} Thickness (varied \SI{0.4}{}--\SI{50}{nm}; $1$--$200$ unit cells). \textbf{c} Convergence angle and \textbf{d} tilt variations. Locally enhanced contrast from a red box is in inset in b. \label{fig4}}
\end{figure}

We examine further the imaging properties of Symmetry STEM by calculating a line scan over \ce{Ce-B-Ce} atomic columns in $\left<100\right>$ \ce{CeB6} for different defocus and probe size and sample thickness and tilt (Fig. \ref{fig4}). The symmetry element is chosen to be a $180^{\circ}$ rotation and the base parameters are as specified in Methods - STEM simulations and kept constant unless otherwise noted. For comparison, we have also calculated the corresponding BF, ABF and ADF STEM line scans (Supplementary Fig. S1).

\textit{Dependence on Thickness and Defocus:} As S-STEM probes local symmetry within the sample, the contrast does not change at the centre of atomic sites due to the change of defocus nor thickness (Figs. \ref{fig4}a and b respectively). This is a consequence of the fact that, while the scattered intensity distribution can vary rapidly with thickness, accelerating voltage and defocus, its' symmetry remains invariant. This is a great advantage over standard methods for which the sign and magnitude of the signal can change. For all thicknesses, the intensity peaks at atomic sites are an almost constant, extremely narrow width ($\sim$\SI{0.25}{\angstrom}), with some local variations in the vicinity of the \ce{B} octahedra likely due to scattering onto nearby high symmetry sites. Nevertheless, even in the presence of this ``cross-talk'', the narrow peak persists (inset Fig. \ref{fig4}b).

\textit{Resolution and Probe size:} Symmetry-STEM is an example of a “two-step” imaging system (as defined by Gureyev and Paganin et al in Refs. \cite{gureyev2019spatial,paganin2019spatial}) with resolution a consequence of both the experimental system and the virtual post-processing  (see Supplementary information). The sharpness of an intensity maxima and consequent ability to resolve two features in S-STEM depends on the ability of the probe to detect a symmetry change. In other words, it depends on the \textit{probe size} relative to the \textit{rate of change} with position of the symmetry of the local specimen potential. The resolution of Symmetry STEM images is therefore only a function of the direct electron-optical imaging system in so far as this system defines a probe size (which is set by the collective effect of the spatial coherence function, probe-forming aberrations and aperture size \cite{maunders2011practical}).

\begin{figure}[t]
	\includegraphics[width=\columnwidth]{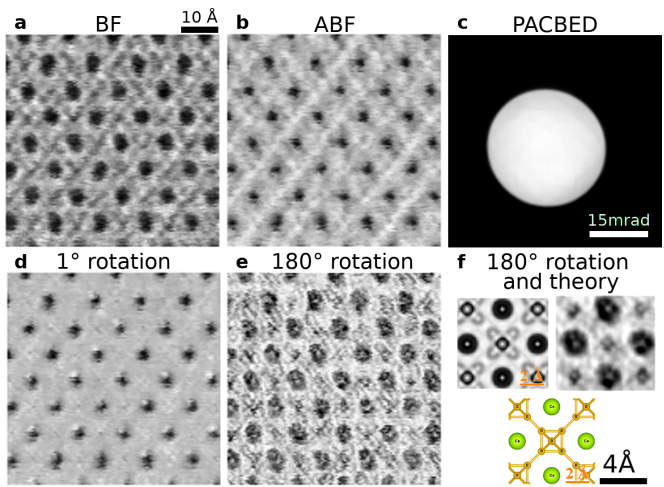}
	\caption{Experimental S-STEM imaging. 128x128 scan dataset collected on FEI Titan$^3$ 80--300 FEGTEM equipped with a pixelated EMPAD detector \cite{tate2016high}. The probe convergence semi-angle was \SI{15}{mrad}. \textbf{a} Reconstruction of BF signal from $0$--\SI{7.5}{mrad}. \textbf{b} ABF signal reconstructed from annulus $7.5$--\SI{15}{mrad}. \textbf{c} averaged diffraction pattern for the whole sample. \textbf{d} $1^{\circ}$ rotation S-STEM image. \textbf{e} $180^{\circ}$ rotation S-STEM image. \textbf{f} Comparison of simulation and averaged experiment for S-STEM for $180^{\circ}$ degree rotation.\label{fig5}}
\end{figure}

In the aberration-free, spatially coherent calculations, the convergence semi-angle defines the probe size and was varied from $0.5$--\SI{40}{mrad} (Fig. \ref{fig4}c). The signal is constant and identity at \SI{<4}{mrad} (\SI{<2.4}{\angstrom} probe FWHM) because the probe is greater than the unit cell in this perfect crystal,  so no change in symmetry can be detected. Put another way, the CBED patterns have non-overlapping CBED discs and hence do not change with position.

From the calculations, it can be seen that when the convergence semi-angle (\SI{>4}{mrad}) generates a probe FWHM (\SI{<2.4}{\angstrom}) comparable to the atomic radius ($\sim$\SI{2.35}{\angstrom} van de Waals) we start to resolve clearly the two \ce{Ce} columns. With higher convergence ($\sim$\SI{7}{mrad}), the probe FWHM ($\sim$\SI{1.7}{\angstrom}) approaches two thirds of the atomic radius and the image peaks sharpen significantly to $\sim$\SI{0.4}{\angstrom} FWHM (Fig. \ref{fig4}c, compared with $\sim$\SI{1.3}{\angstrom} FWHM for the corresponding image peaks in BF-, ABF\mbox{-}, ADF-STEM images in Supplementary Fig. S2b,c and d). This trend continues to $\sim$\SI{22}{mrad}, where the probe FWHM is \SI{0.44}{\angstrom}, less than a quarter of the atomic radius. Here the image peak width is $<$\SI{0.2}{\angstrom}, which is half of the diffraction-limited probe size set by this convergence angle and a fifth of the image peak width for conventional BF-, \mbox{ABF-} and ADF-STEM modes (Supplementary Fig. S2b,c and d). The sharpness of the peak reduces for convergence angles larger than $\sim$\SI{25}{mrad}, possibly due to scattering onto nearby atomic sites promoted by the larger transverse momentum of the incident probe.

\textit{Dependence on Tilt:} Tilt of a sample is a crucial parameter to study as  it changes the excitation errors and hence the symmetry of the scattering matrix and resulting CBED pattern. The sample was tilted from the $\left<100\right>$ zone axis by $-1^\circ$ to $3^\circ$ in the plane of the line scan (Fig. \ref{fig4}d). When the sample is tilted more than $\sim 0.05^\circ$ the symmetry peak related to \ce{Ce} starts to disappear, while the \ce{B} peak broadens but persists (likely due to the \ce{B} octahedra acting effectively as a single scatterer). In some crystalline specimens, this sensitivity to tilt could be helpful for the precise alignment of the crystal along a zone axis, without the problem of significant defocus change with tilt (Fig. \ref{fig4}a). 

\subsection*{Experiment} Finally, we compare the Symmetry STEM analysis of simulated data with experimental data from \ce{CeB6} (Fig. \ref{fig5}). The scanned CBED data was collected at \SI{300}{kV} on an early generation double-spherical aberration corrected FEI Titan$^3$ 80--300 FEGTEM equipped with a pixelated EMPAD detector \cite{tate2016high}. Aberrations were largely corrected within the convergence semi-angle of \SI{15}{mrad}. Conventional BF and ABF STEM images were reconstructed (Fig. \ref{fig5}a and b, respectively) to compare with the Symmetry STEM signals with the same symmetry operations as applied in Fig. \ref{fig3} (Fig. \ref{fig5}). (The HAADF signal was not collected here because the large angular field of view required would constrain the symmetry measurement from the central disk area.) The $1^\circ$ and $180^\circ$ rotation images (Figs.  \ref{fig5}d and e, respectively), show contrast closely related to the theoretical calculations (Fig. \ref{fig3}). In particular, the $180^\circ$ rotation image shows extremely sharp peaks surrounded by dark `moats', corresponding to the symmetry maxima when the probe is positioned at the absolute centre of the \ce{Ce} atomic column and the break of symmetry as soon as the beam shifts slightly from the centre but remains on the column, as seen in the calculations (Fig. \ref{fig3}f). We can also see the symmetric position of the central \ce{B} atomic column, however the asymmetric sites of other \ce{B} atomic positions are not as clear, most likely due to imperfect instrument and specimen stability and a lack of local $180^\circ$ rotation symmetry in the position. The $1^\circ$ rotation image shows very strong signal to noise at the \ce{Ce} columns and also shows some residual specimen tilt effects. An average unit cell image (from five unit cells) is compared with the calculated image in Fig. \ref{fig5}f.

\section*{Discussion} 

In Symmetry STEM, the symmetry of the local specimen potential defines the mathematical symmetry of the dynamical N-dimensional scattering matrix \cite{cowley2006electron,buxton1976symmetry,tanaka2010point} and this, in turn, defines the symmetry of the scattered intensity distribution (CBED pattern) which is extracted using Equation \ref{eq:image_contrast}, providing an intrinsically different image contrast mechanism. 

For perfect crystals, a reduction in symmetry can be detected as soon as an atomic-scale probe shifts a few picometres from the centre of an atomic column. Our calculations show this can generate extremely sharp image peaks at atomic column positions (Fig. \ref{fig3}i) that have smaller FWHM than the corresponding peaks/troughs in conventional diffraction-limited BF-, ABF- and ADF-STEM images derived from the same 4D-STEM data set (see Figs. \ref{fig4}i and Supplementary S4).

For unknown structures, the presence of a mirror or any point symmetry element can be detected through the automated application of the Symmetry STEM algorithm (Equation \ref{eq:image_contrast}) for any rotation angle, as illustrated in Figure \ref{fig2} and the Movies S1 and S2. Similarly, this can enable defects, which break local symmetry, to be highlighted.

In experimental data, the practical limit is the stability of the instrument and specimen, collectively ‘scan noise’. The proof-of-concept examples in Fig. \ref{fig5} were taken on a decade-old instrument. The on-going improvements in instrument stability and detector speeds bode well for the further development of this technique. 
The remarkable robustness of S-STEM images to thickness, accelerating voltage and defocus is because the mathematical \textit{symmetry} of the scattering matrix and hence the CBED pattern symmetry does not depend on these quantities. The acute sensitivity to tilt is because it does. The latter provides an opportunity for an automated, high precision tilt alignment \cite{hajirassouliha2013fpga}.
The potential of Symmetry STEM to obtain atomic resolution images of light and heavy atoms, from thick and thin crystals across a wide selection of accelerating voltages opens up a range of applications in material science that are otherwise challenging to image, including thick and beam sensitive specimens. The ability to image defects due to the change in symmetry they induce, also opens new opportunities, including the imaging of dopant atoms and dislocations. The method could also be applied to the imaging of atomic magnetic fields in electromagnetic circular dichroism \cite{rusz2014achieving}. Furthermore, Symmetry STEM is likely to be sensitive to atomic displacements which induce a local change in symmetry, such as can occur with strain or octahedral tilts.

\section*{Methods}
\subsection*{STEM imaging} Imaging was performed using a double- spherical-aberration corrected Titan$^3$ 80-300 FEG-TEM equipped with an  EMPAD pixelated detector\cite{tate2016high} and operating at $300\,kV$ electron energy with a $15\,mrad$ probe convergence semi-angle. The dataset was a 4D-STEM data of 128x128 probe positions with the size 1GB. 

\subsection*{Symmetry analysis} A custom made C++ code using GPU accelerated library ArrayFire \cite{Yalamanchili2015} was used to analyse symmetry in two dimensional diffraction pattern images. A maximum value of a normalised cross(phase)-correlation for a specific symmetry operation is plotted for each point of a scan. 

\subsection*{STEM simulations} Prismatic software package \cite{pryor2017streaming,orphus2017afast} was used to simulate 4D-STEM datasets. The parameters were: \ce{CeB6} sample was \SI{41.3}{nm} (Fig. \ref{fig1}) or \SI{2.8}{nm} (Figs. \ref{fig3} and \ref{fig4}) thick,  in $\left<100\right>$ orientation, convergence semi-angle was \SI{15}{mrad}, acceleration voltage \SI{300}{kV}, slice thickness was $\SI{1}{\angstrom}$, beam was focused on the top surface and no aberrations were used. The size of a single dataset was \SI{2.5}{GB} and included 124x124 probe positions.

\subsection*{Acknowledgements}
We thank Prof. David Paganin for helpful comments
	on the manuscript and A/Prof Matthew Weyland for helpful suggestions for STEM alignment. We thank Dr Ding Peng and A/Prof Philip Nakashima for the tripod polished \ce{CeB6} specimen from a crystal supplied by Mr. P. Hanan via Prof AWS Johnson \cite{peng2017specimen}. This work was supported by the Australian Research Council (ARC) grant DP150104483 and used instruments at the Monash Centre for Electron Microscopy funded by ARC grant LE0454166.


\bibliographystyle{naturemag}
\bibliography{symmetry}



\end{document}